\documentclass[pra,preprint,amsfonts,amssymb,amsmath,titlepage,floatfix,showpacs,superscriptaddress]{revtex4} 

\usepackage{psfrag}
\usepackage{bm}

\usepackage{bm}
\usepackage{graphicx}  
\usepackage{latexsym}                
\usepackage{amsmath}     
\usepackage{amsfonts}  
\usepackage{amssymb}
\usepackage{amsthm}
\usepackage{dcolumn}
\usepackage{color}
\usepackage{comment}

\newcommand{\x}{{\bf x}}

\newcommand{\E}{{\bf E}}
\newcommand{\e}{{\bf e}}

\newcommand{\be}{\begin{equation}}
\newcommand{\ee}{\end{equation}}
\newcommand{\bea}{\begin{eqnarray}}
\newcommand{\eea}{\end{eqnarray}}
\newcommand{\bdm}{\begin{displaymath}}
\newcommand{\edm}{\end{displaymath}}

\newcommand{\cE}{{\cal E}}

\def\pder#1#2{\frac{\partial #1}{\partial #2}}
\def\lsim{\raise0.3ex\hbox{$<$\kern-0.75em\raise-1.1ex\hbox{$\sim$}}}

\def\pder#1#2{\frac{\partial #1}{\partial #2}}
\def\softt{{\leavevmode\setbox1=\hbox{t}%
\hbox to \wd1{t\kern-0.6ex{\char039}\hss}}}

\begin{document}
\title{Correlated Directional Atomic Clouds via Four Heterowave Mixing}

\date{\today}
\author{L.~F.~Buchmann}
\affiliation{Institute of Electronic Structure and Laser, FORTH, P.O. Box 1527, Heraklion 71110, Crete, Greece}
\affiliation{Department of Physics, University of Crete, P.O. Box 2208, Heraklion 71003, Crete, Greece}
\author{G.~M.~Nikolopoulos}
\affiliation{Institute of Electronic Structure and Laser, FORTH, P.O. Box 1527, Heraklion 71110, Crete, Greece}
\author{O.~Zobay}
\affiliation{School of Mathematics,
University of Bristol,
University Walk, Bristol, BS8 1TW, UK}
\author{P.~Lambropoulos}
\affiliation{Institute of Electronic Structure and Laser, FORTH, P.O. Box 1527, Heraklion 71110, Crete, Greece}
\affiliation{Department of Physics, University of Crete, P.O. Box 2208, Heraklion 71003, Crete, Greece}

\begin{abstract}
We investigate the coherence properties of pairs of counter-propagating 
atomic clouds, produced in superradiant Rayleigh scattering off atomic condensates.
It is shown that these clouds exhibit long-range spatial coherence and 
strong nonclassical density cross-correlations, which make this scheme 
a promising candidate for the production of highly directional 
nonclassically correlated atomic pulses. 
\end{abstract}
 	 

\maketitle
\section{Introduction}Atom-atom correlations, and the engineering of ``nonclassical'' states of 
atoms that exhibit quantum correlations, are currently attracting intense 
theoretical and experimental interest in the framework of ultracold quantum gases. 
In most of these studies, Bose-Einstein condensation plays a central role, 
since correlated atomic beams can be generated by manipulating appropriately 
preformed condensates \cite{MoleculeBEC,PuMey00,Camp06,Aspect-co,OgrKhe09,DeuDru07,VogXuKett00}.  
Directionality of the produced atom pairs, is of vital 
importance for many potential applications (e.g., subshot noise precision measurements,
tests of quantum nonlocality, and presumably quantum information processing) but, 
as discussed in \cite{VarMoo02}, it is difficult to be achieved within existing schemes  without seeding (e.g., see \cite{VogXuKett00}).

Superradiant Rayleigh scattering from an elongated atomic condensate \cite{InoChiSta99}, 
has the potential to produce highly directional counter-propagating matter waves, 
which exhibit correlations as shown in \cite{Mey-and-co}. Moreover, 
in contrast to other proposals, the two matter waves have well defined spatial profiles, 
and one can tune their macroscopic populations. 
So far, however,  the usefulness of the scheme remains debatable, since 
there has been no theoretical quantitative analysis of the coherence properties of 
the matter waves, and the type of correlations involved.

Most of these issues are addressed in this article, within a model that 
treats the scattered photons and the matter waves quantum mechanically.  
Our model is expected to describe accurately the essential aspects of the process. 
Moreover, it includes spatial propagation effects, which are crucial for the 
thorough understanding of condensate superradiance \cite{ZobNikPRA}, 
and were not included in \cite{Mey-and-co,You-and-co}.

It is shown that the counter-propagating atomic clouds exhibit long-range spatial coherence.
Their  density cross-correlations are enhanced due to the formation of correlated 
atom bunches, which are also responsible for violation of the classical Cauchy-Schwarz 
inequality, relative number squeezing and entanglement.
Unlike earlier schemes \cite{Camp06,Aspect-co,OgrKhe09,DeuDru07,VogXuKett00}, we study correlations in the context of mixing two optical-- and two matter waves, hence the name ``Four Heterowave Mixing''.  The present results shed light on the physics of such a mixing process, 
and they determine the operational regimes for possible applications and future experiments. 

\section{Model}
The system involves an elongated condensate of length $L$, oriented along the $z$ axis and consisting of 
$N$ atoms \cite{InoChiSta99}.  
It is exposed to a linearly polarized pump laser pulse  
$\E_l(\x,t)=\cE_l(t){\bf e}_y [e^{{\rm i}(k_lx-\omega_l t)}+c.c.]/2$, 
with $\omega_l = ck_l$, traveling in the $x$ direction. 
The laser is assumed to be far off-resonant 
from any atomic transition, inducing thus Rayleigh scattering.

Due to the coherent nature of the condensate,
successive Rayleigh scattering events are strongly correlated, leading to collective 
superradiant behaviour \cite{InoChiSta99,Mey-and-co,You-and-co,ZobNikPRA}. 
Moreover, as a result of the cigar shape of the 
condensate, the gain is largest when the scattered photons leave the 
condensate along its long axis, in the so called endfire 
modes with momenta ${\bf k} \approx \pm k_l{\bf e}_z$ and frequency $\omega\approx \omega_l$.
As a consequence, the recoiling atoms have well-defined momenta and appear 
in distinct atomic side modes.  In the side mode $(n,m)$, 
atoms possess momentum ${\bf q}\approx \hbar k_l(n\e_x + m\e_z)$, while the corresponding 
frequency is approximately given by $\omega_{n,m} \approx (n^2+m^2)\omega_r$, where 
$\omega_r = \hbar k_l^2/2M$ is the recoil frequency, and $M$ the atomic mass.  
In this notation, the ``side mode" $(0,0)$ describes the condensate at rest, while 
$(1,\pm 1)$ and $(-1,\pm 1)$ are the first-order forward and backward atomic 
side modes, respectively. 

We are here concerned with the early stage of the process,
where only first-order atomic side modes become significantly populated. The condensate 
remains practically undepleted and can be treated as a time-independent 
classical field $\Phi({\bf x})=\varphi(z)\varphi_\perp(x,y)$, 
where $\varphi(z)$ and $\varphi_\perp(x,y)$ are the longitudinal and 
transverse wave functions. 
For short times, the coupling between the counter propagating optical 
end-fire modes can be neglected. Taking advantage of the symmetry of the system with respect to the $x$-axis,
we focus on the dynamics of the single endfire mode (taken to be 
monochromatic with ${\bf k}= +k_l{\bf e}_z$), and 
the first-order side modes coupled by it [i.e., $(1,-1)\equiv (+)$ and 
$(-1,1)\equiv (-)$].  

Under the slowly-varying-envelope approximation (SVEA), we decompose the operators for 
the matter wave and the positive-frequency electric fields as 
$\hat{\Psi}(\x,t) =\Phi(\x) +\varphi_\perp
\sum_{j}\hat\psi_j(z,t) e^{-{\rm i}[\omega_jt-j(k_lx -kz)]}$, and 
$\hat{\E}^{(+)}(\x,t) = \E_l^{(+)}(\x,t)
+\sqrt{\frac{\hbar\omega}{2\varepsilon_0}}u_\perp \hat e(z,t)\e_y e^{-{\rm i}(\omega t- kz)}$,
with $u_\perp(x,y)$ the transverse endfire-mode profile, and $j\in\{+,-\}$.
The operators $\hat\psi_j(z,t)$, $\hat e(z,t)$ are annihilation 
operators for side-- and endfire-mode fields with transverse behavior fixed 
by $\varphi_\perp$ and $u_\perp$ (see appendix). 
By construction, these operators obey standard commutation relations with, however, 
Dirac delta functions having finite width of about $1/k$ (see appendix). 
Inserting the SVEA expansions into the Maxwell-Schr\"odinger equations for the coupled 
matter-wave and electric fields 
\cite{GroHar82}, one can derive equations of motion 
for the operators $\hat\psi_j(z,t)$ and $\hat e(z,t)$ 
\cite{longpaper}.  
Rescaling to dimensionless time $\tau=2\omega_r t$ 
and length $\xi=k_lz$, we obtain
\begin{subequations}\label{eom}
\bea
&&\hspace*{-0.5cm}\textrm{i}  \pder {}{\tau}\hat\psi_{+}^\dag(\xi,\tau) = -\kappa  \hat e(\xi,\tau) \varphi^*(\xi),\label{eom1-1} \\
&&\hspace*{-0.5cm}\textrm{i}  \pder {}{\tau} \hat\psi_{-}'(\xi,\tau) = \kappa\hat e(\xi,t)\varphi(\xi)+2\hat{\psi}_{-}^\prime(\xi,\tau),\label{eom-11}\\
&&\hspace*{-0.5cm}\textrm{i} \left(\pder{}\tau + \chi \pder{}\xi\right)\hat e(\xi,\tau)=
\kappa \left[ \varphi^*(\xi)\hat{\psi}_{-}'(\xi,\tau)\right. \nonumber\\
&&\hspace*{4.0cm}\left.+\hat{\psi}_{+}^\dag(\xi,\tau)\varphi(\xi)\right]\label{eome+}
\eea
\end{subequations}
with $\hat{\psi}_{-}'=\hat{\psi}_{-}e^{-2{\rm i}\tau}$, $\chi=ck_l/(2\omega_r)$
 and $\kappa=g\sqrt{k_lL }/(2\omega_r)$ \cite{remark1}.
The atom-photon coupling is given by
$g=|{\bf d}|^2\mathcal{E}_l/(2\hbar^2 \delta) \sqrt{\hbar \omega/(2\varepsilon_0L)}\int dxdy\, \varphi_\perp^2 u_\perp$, 
where ${\bf d}$ is the atomic dipole moment, and $\delta$ the detuning of the laser from the nearest atomic transition. By discarding backwards recoiling atomic modes in Eqs. (\ref{eom}), the remaining equations describe conventional superradiance.

Solutions to the system of Eqs. (\ref{eom}) can be 
expressed as integrals involving the operators evaluated at the boundary 
of their domain --- i.e. at $\xi=0$ and $\tau>0$ or vice versa at $\xi>0$ 
and $\tau=0$ --- by applying Laplace transform techniques. Due to the 
large value of $\chi$, retardation effects can be neglected, 
and the solutions for the matter wave operators read \cite{longpaper}
\begin{subequations}\label{sols}
\bea
\hat{\psi}_{+}^\dag(\xi,\tau)&=&\hat{\psi}_{+}^\dag(\xi,0)
\nonumber\\
&+&{\rm i} \kappa \varphi^*(\xi)
\int_0^\tau d\tau^\prime \hat{e}(0,\tau^\prime ) F_{1,0}(\gamma_{\xi,0},\tau-\tau^\prime)\nonumber\\
&+&\Gamma\varphi^*(\xi)\bigg\{\int_{0}^{\xi}d\xi^\prime
\varphi(\xi^\prime)
\hat{\psi}_{+}^\dag(\xi^\prime,0) 
F_{2,0}(\gamma_{\xi,\xi'},\tau)\nonumber\\
&+&\int_{0}^{\xi}d\xi^\prime
\varphi^*(\xi^\prime)
\hat{\psi}_{-}^{\prime}(\xi^\prime,0)F_{1,1}(\gamma_{\xi,\xi'},\tau)\bigg\},
\label{psi_f1}
\\
\hat{\psi}_{-}^\prime(\xi,\tau)
&=&e^{-{\rm i}2\tau}\hat{\psi}_{-}^\prime(\xi,0)\nonumber\\
&-&{\rm i}\kappa\varphi(\xi)
\int_0^\tau d\tau^\prime \hat{e}(0,\tau^\prime ) F_{0,1}(\gamma_{\xi,0},\tau-\tau^\prime)
\nonumber\\
&-&\Gamma\varphi(\xi)\bigg\{
\int_{0}^{\xi}d\xi^\prime
\varphi(\xi^\prime)
\hat{\psi}_{+}^\dag(\xi^\prime,0) F_{1,1}(\gamma_{\xi,\xi'},\tau)\nonumber\\
&+&
\int_{0}^{\xi}d\xi^\prime
\varphi^*(\xi^\prime)
\hat{\psi}_{-}^\prime(\xi^\prime,0) F_{0,2}(\gamma_{\xi,\xi'},\tau)\bigg\}
\label{psi_b1}.
\eea
\end{subequations}
We have introduced $\Gamma=\kappa^2/\chi$ and the functions
$F_{\mu,\nu}(\alpha,\beta)={\mathcal L}_{p\to \beta}^{-1}\left\{
e^{\alpha/p}e^{-\alpha/(p+2{\mathrm i})} p^{-\mu}(p+2{\mathrm i})^{-\nu}\right\}$,
 with ${\mathcal  L}^{-1}$ denoting the inverse Laplace 
transform, while $\gamma_{\xi,\xi'}=\Gamma\int_{\xi'}^\xi d\zeta |\varphi(\zeta)|^2$. 
The functions $F_{\mu,\nu}$ can be expressed explicitly as
combinations of Bessel functions and integrals thereof (see appendix). 

Experimental observations of superradiance from condensates distinguish between 
the so-called weak- and strong-pulse regimes \cite{InoChiSta99}. 
Typically, the former (WP) regime is characterized by coupling
constants $g\sim10^5{\rm s}^{-1}$  and $\Gamma N\simeq 1$, while for the
latter (SP) regime $g\sim10^6{\rm s}^{-1}$  and $\Gamma N\gg 1$. 
Throughout our simulations, 
we chose $\Gamma N=1$ and $\Gamma N=100$ for the two regimes. 
The condensate at rest was taken to be Thomas-Fermi distributed, 
i.e. $\varphi(z)=\sqrt{6(L z - z^2)\Theta(z)/L^3}$, with $L=130\mu{\mathrm m}$ and 
$\Theta(\cdot)$ the Heaviside step function; 
containing $N=10^6$ $^{87}$Rb atoms. The incoming laser pulse was chosen 
to have a rectangular profile and $k_l=8.05\times 10^6 {\mathrm m}^{-1}$.

\section{Coherence and Correlations} The first-order correlation function of the side mode $(j)$  
is $G_{jj}^{(1)}(\xi_1,\xi_2;\tau)=\langle\hat{\psi}_j^\dag(\xi_1,\tau)\hat{\psi}_j(\xi_2,\tau)\rangle$. 
It characterizes the coherence properties of the side mode, 
and quantifies the characteristic coherence length over which 
phase correlations exist. Using Eqs. (\ref{sols}), the commutation relations for the 
fields \cite{remark3}, and applying the boundary values 
$\langle\hat{e}^\dag\hat{e}(\xi=0,\tau)\rangle=\langle\hat{\psi}_{j}^\dag\hat{\psi}_{j}(\xi,\tau=0)\rangle=0$ 
one can easily obtain analytic expressions for $G_{jj}^{(1)}(\xi_1,\xi_2;\tau)$ for any time $\tau$.  
In practice, however, it might not be feasible to resolve the exact atomic 
positions $\xi_1$, and $\xi_2$ in the two sidemodes. For direct comparison to possible 
experiments, we consider the volume averaged degree of coherence, defined as \cite{Glauber99} 
\[\tilde{g}_{jj}^{(1)}(\Delta\xi;\tau) =\frac{
\int_{0}^\Lambda d\xi_1 |G_{jj}^{(1)}(\xi_1,\xi_1+\Delta\xi;\tau)|}{\int_{0}^\Lambda d\xi_1 
\sqrt{{\cal N}_{jj}(\xi_1,\Delta\xi;\tau)}
},
\] 
where ${\cal N}_{jj^\prime}(\xi_1,\Delta\xi;\tau)=\langle \hat{n}_{j}(\xi_1,\tau)\rangle\langle \hat{n}_{j^\prime}(\xi_1+\Delta\xi,\tau)\rangle$, with $\langle \hat{n}_j(\xi,\tau)\rangle=\langle\hat{\psi}_j^\dag(\xi,\tau)\hat{\psi}_j(\xi,\tau)\rangle$, $\Delta\xi=\xi_2-\xi_1$  and $\Lambda\equiv k_lL$. 

\begin{figure}
\includegraphics[scale=0.5]{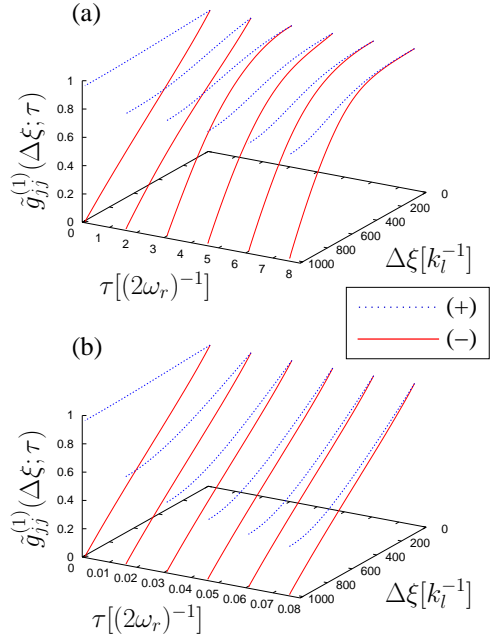}
\caption{
Degree of first-order coherence for the $j$th side-mode at various times, in 
the WP (a) and SP (b) regimes. 
The function is symmetric with respect to $\Delta\xi=0$.
}\label{MWg1}
\end{figure}

As depicted in Fig. \ref{MWg1}, $\tilde{g}_{jj}^{(1)}(\Delta\xi;\tau)$ 
attains its maximum for $\Delta\xi=0$, where $\tilde{g}_{jj}^{(1)}(0;\tau)=1$.
Let the coherence length of the side mode $(j)$ be 
the spatial separation $\lambda_j$ for which, 
$\tilde{g}_{jj}^{(1)}(\lambda_j;\tau)\approx 1/e$ \cite{CohLength}.  
The matter-waves associated with the two side modes can be considered 
as first-order-coherent for any two points separated by 
$|\Delta\xi|<\lambda_{j}$.
According to  Fig. \ref{MWg1}, the long-range order of the condensate is transferred to the atoms in the side modes $(\pm)$. 
More precisely, for relatively short times, the coherence length of the side mode 
$(+)$ is $\lambda_1\approx \Lambda$, while for the 
side mode $(-)$ it is somewhat smaller $\lambda_2\approx 0.6\Lambda$.
This is because backward recoiling atoms 
are created when incoherent light of the endfire mode is scattered by 
condensed atoms (as opposed to forward recoiling atoms, which involve scattering 
of coherent laser light), and thus the transfer of coherence is not as 
efficient as for the side mode $(+)$.
In the SP regime, the coherence length of the side mode $(+)$ 
decreases as time goes on, approaching the one of 
the backward side mode, which remains practically constant. 
In the WP regime, however, the corresponding drop is not so prominent, 
and instead we observe a growth of 
the spatial coherence for the side mode $(-)$.

Density correlations are described through the  second-order 
correlation functions 
$G^{(2)}_{jj^\prime}(\xi_1,\xi_{2};\tau)=\langle:\hat{n}_{j}(\xi_1,\tau)\hat{n}_{j^\prime}(\xi_{2},\tau):\rangle$. 
This quantity reflects the probability of finding an atom of type $(j^\prime)$ at a position $\xi_{2}$, 
given that an atom of type $(j)$ has been detected at position $\xi_{1}$.  
Using Eqs. (\ref{sols}), one can express $G^{(2)}_{jj^\prime}(\xi_1,\xi_{2};\tau)$ 
in terms of first-order correlation functions. 
The corresponding volume averaged normalized correlation functions can be 
defined in analogy to $\tilde{g}_{jj}^{(1)}$ \cite{Glauber99}, obtaining  
\bea
\label{Acorrfunc}
\tilde{g}_{jj^\prime}^{(2)}(\Delta\xi;\tau) &=&1+
\frac{\int_{0}^\Lambda d\xi_1 |\rho_{jj^\prime}(\xi_1,\xi_1+\Delta\xi;\tau)|^2}{\int_{0}^\Lambda d\xi_1 
{\cal N}_{jj^\prime}(\xi_1,\Delta\xi;\tau)},
\eea
for $\rho_{jj}=G_{jj}^{(1)}(\xi_1,\xi_2;\tau)$ and 
$\rho_{+-} =\langle\hat{\psi}_+(\xi_1,\tau)\hat{\psi}_-(\xi_2,\tau)\rangle$.

In view of Eq. (\ref{Acorrfunc}), the function  
$G_{jj}^{(1)}(\xi_1,\xi_2;\tau)$ determines to a large extent 
the density autocorrelation function $\tilde{g}_{jj}^{(2)}(\Delta\xi;\tau)$. 
The behavior of $\tilde{g}_{jj}^{(2)}(\Delta\xi;\tau)$ throughout the evolution 
of the system resembles the behavior of the first-order coherence, 
albeit with somewhat different profiles, not shown here. 
In contrast to $\tilde{g}_{jj}^{(1)}(\Delta\xi;\tau)$, however, 
 $\tilde{g}_{jj}^{(2)}(0;\tau)\to 2$, as $|\Delta\xi|\to 0$. This is a manifestation 
of correlations between atomic densities at two different points separated by 
$|\Delta\xi|\ll\lambda_j$. In other words, the recoiling atoms in the side 
mode $(j)$ tend to appear bunched and thus,  
detecting an atom in position $\xi_1$, significantly increases  
the probability of detecting another atom close to it. The spatial extent 
of atom bunches is of the order of $\lambda_j$.  
For $|\Delta\xi| >\lambda_j$, $\tilde{g}_{jj}^{(2)}(0;\tau)\to 1$, 
indicating that atomic densities become uncorrelated for larger separations.  

\begin{figure}
\includegraphics[scale=0.475]{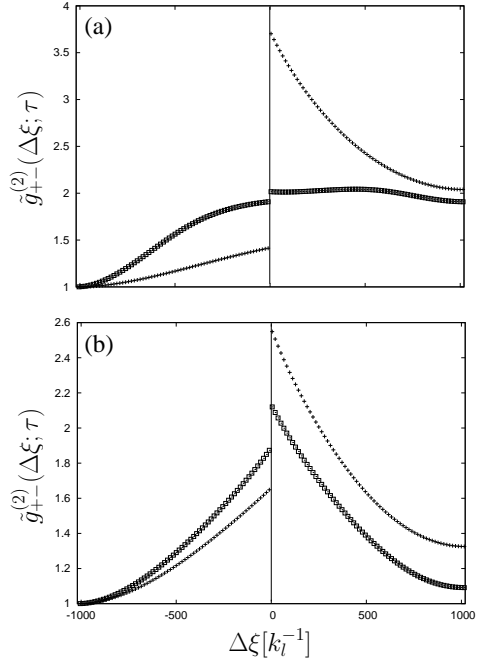}
\caption{
Normalized density cross-correlation function at two different times. 
(a) WP regime: $\tau=1.025$ $(+)$, $\tau=4.025$ $(\Box)$; 
(b) SP regime: $\tau=0.03025$ $(+)$, $\tau=0.12075$ $(\Box)$.
}\label{MWcorr}
\end{figure}

Density cross-correlations between the two side modes, 
are described through $\tilde{g}_{+-}^{(2)}(\Delta\xi,\tau$), 
which is plotted in Fig. \ref{MWcorr}. 
For any time $\tau$, there is a discontinuity at $\xi_1=\xi_2$ due to 
one of the terms in  $\rho_{+-}$, which involves $\Theta(\xi_1-\xi_2)$.
The physical reason for this discontinuity 
is the fact that an atom at  $\xi_1$ in the sidemode ($+$) can be correlated to 
another atom at $\xi_2$ in the sidemode ($-$), only when the latter 
is created by scattering the endfire photon that was emitted by the former. 
This process (to be referred to as photon exchange hereafter) 
is only possible for $\xi_2\ge\xi_1$ since endfire photons 
possess momenta $+k{\bf e_z}$. 

The enhanced correlations observed in Fig. \ref{MWcorr} for $|\Delta\xi|<\lambda_+$, 
arise through the interplay between atom bunching, exhibited by the side mode $(+)$, 
and photon exchange. A recoiling atom of type $(-)$, 
is not correlated only to the atom of type $(+)$ that has emitted the 
photon, but rather to the entire bunch. Moreover, due  
to bosonic enhancement, the production of a backward-recoiling  
atom at a given position, stimulates the creation of additional atoms of the 
same type nearby. 
We have thus the production of correlated atomic bunches of type $(+)$ and $(-)$.
This is clear in Fig. \ref{MWcorr}, where the cross-correlation function is 
always larger than unity, and increases with decreasing $|\Delta\xi|$, attaining 
its maximum value for $\Delta\xi\to 0^\pm$. Thus, measuring an atom 
in one mode significantly increases the probability of measuring one in the 
counter-propagating mode. The spatial dependence $g_{+-}^{(2)}(\Delta\xi,\tau)$ 
at various times, bears analogies to the corresponding behavior of 
$g_{jj}^{(1)}(\Delta\xi,\tau)$. 
As a general remark, in the SP regime, we find a much faster decay of 
$g_{+-}^{(2)}(\Delta\xi,\tau)$ compared to the WP regime.

\begin{figure}
\includegraphics[scale=0.7]{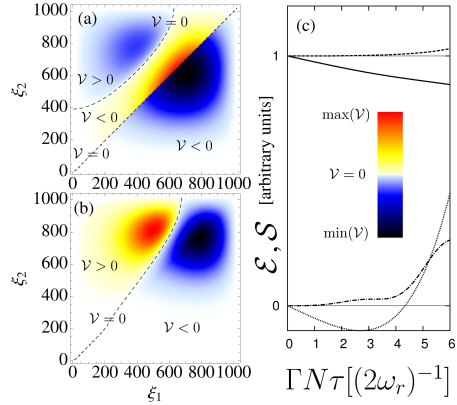}
\caption{(Color Online) (a,b) Density plots of ${\mathcal V}$ at $\Gamma N\tau\simeq 3$.
(c) Time evolution of ${\cal S}$ (upper) and ${\cal E}$ (lower).
WP regime: (b), dashed and dot-dashed lines; SP regime: (a), solid and dotted lines.} 
 \label{figureviol}
\end{figure}

A further, qualitative analysis of correlations can be obtained by
means of inequalities. It is known that ``classical'' fields 
satisfy the inequality ${\mathcal V}(\xi_1,\xi_2;\tau)\leq 0$ with 
$
{\mathcal V}=[G^{(2)}_{+-}(\xi_1,\xi_2;\tau)]^2-G^{(2)}_{++}(\xi_1,\xi_1;\tau)G^{(2)}_{--}(\xi_2,\xi_2;\tau)
$,
for all times $\tau$ and pairs $(\xi_1,\xi_2)$ \cite{Glauber99}. 
A quantum field, however, can violate this inequality if its P-representation 
attains negative values in a certain region of space and/or time. 
As depicted in Figs. \ref{figureviol}(a,b), 
in our system, photon exchange may lead to violation of this inequality 
only for $\xi_2\gtrsim\xi_1$. Such a violation can be also 
associated with squeezing or entanglement. 
The squeezing in the population difference between the two side modes 
is quantified by, 
$\mathcal{S}(\tau)=1+\langle{\hat{R}_+}\rangle^{-1}\langle:(\Delta \hat{R}_-)^2:\rangle$, 
where $\Delta\hat{O}=\hat{O}-\langle\hat{O}\rangle$, 
$\hat{R}_\pm(\tau)=\hat{N}_+(\tau)\pm\hat{N}_-(\tau)$ and 
$\hat{N}_j(\tau)=\int_0^\Lambda d\xi \hat{n}_j(\xi,\tau)$ \cite{MoleculeBEC}. 
As depicted in Fig. \ref{figureviol}(c), the system exhibits squeezing in the 
SP regime only where $\mathcal{S}(\tau)<1$. 
This never happens in the WP regime where 
$\langle:(\Delta \hat{R}_-)^2:\rangle\approx \int_0^\Lambda
\int_0^\Lambda d\xi_1 d\xi_2 |G_{++}^{(1)}(\xi_1,\xi_2;\tau)|^2$. 
A necessary condition for separability of the bipartite system consisting of the 
modes $(\pm)$ is ${\cal E} \geq 0$, where  
${\cal E}(\tau)=\textrm{var}(\hat{A}_++\hat{A}_-)+\textrm{var}(\hat{B}_+-\hat{B}_-)-C_+ -C_-$, 
and $C_j=|\langle [\hat{A}_j, \hat{B}_j] \rangle|$ \cite{Ray-etal}.
We have checked this inequality for 
$A_j(\tau)=\int_0^\Lambda d\xi (\psi_j + \psi_j^\dag)$ and 
$B_j(\tau)={\rm i}\int_0^\Lambda d\xi (\psi_j - \psi_j^\dag)$, which are 
the quadratures of the side mode $(j)$, with respect to its mean momentum.
As depicted in Fig. \ref{figureviol}(c), the two side modes are indeed 
entangled in the SP regime for $0<\Gamma N\tau\lesssim 4.2$. 
When ${\cal E}\geq 0$, the above criterion 
does not allow us to infer anything about the entanglement in the system. 
In view of these results, the SP regime seems to be more interesting for 
practical purposes than the WP regime.

\section{Conclusions} Superradiance from condensates is a promising 
technique for producing atomic clouds, with well defined momenta, 
spatial profiles, as well as tunable macroscopic populations. 
The counter-propagating matter waves typically produced in the 
strong-pulse regime, exhibit long-range coherence 
and nonclassical correlations, which can be explored for various applications.
Our description does not take multi-modal superradiant emission  
into account \cite{InoChiSta99,Mey-and-co}, which can be strongly 
suppressed by reducing the aspect ratio of the condensate \cite{InoChiSta99,Mey-and-co,GroHar82}. The main predictions of our 
model are expected to be valid in more elaborate three-dimensional 
simulations, which can take into account also the propagation of 
atoms. In that case, however, the discontinuous gap in 
$\tilde{g}_{+-}^{(2)}(\Delta\xi;\tau)$ 
is expected to appear as a sharp but continuous increase. 
The experimental techniques that have been developed over recent years allow for 
the direct measurement of correlation functions \cite{Aspect-co,Ritt07}, and should 
be applicable to the verification of the present theoretical predictions. Finally, some of the present results might also apply to other systems which exhibit similarities to superradiance off condensates such as the collective atomic recoil laser \cite{Cola}.

The work was supported by the EC RTN EMALI.

\begin{appendix}{Appendix}

\section{Operator Expansion}
Our operators are expanded as 
\bea
\hat{\psi}_{j}(z,t)&=&\frac{e^{{\mathrm i}\omega_{j}t}}{\sqrt{L}}\sum_{p\in\Delta_{0}}e^{i p z} {\hat c}_{-j k+p}(t),\label{psipm}\\
\hat{e}(z,t)&=&\frac {e^{i\omega t}}{\sqrt L}\sum_{p\in\Delta_{0}} e^{ipz} \hat a_{k+p}(t),
\eea
where $j\in\{-1,+1\}$, $\omega_{\pm 1}=\frac{k^2+k_l^2}{2M}$ and $\Delta_0$  is an interval around zero in momentum space which cannot be chosen larger than $(-k/2,k/2)$ for $\left[\psi_{+}(z,0),\psi_{-}^\dag(z',0)\right]=0$ to hold. Therefore the equal field commutator  $\left[\psi_{j}(z,0),\psi_{j}^\dag(z',0)\right]$ is not a Dirac delta function, but rather a distribution with width of the order $1/k$. 

\section{Inverse Laplace Transforms}
\begin{subequations}
\bea
F_{1,0}(y,z)&=&
I_0\left ( 2\sqrt{y z} \right )\Theta( z)
-\Theta( z)\sqrt{y}\int_0^{ z} dz^\prime \frac{e^{-{\rm
i}2z^\prime}}{\sqrt{z^\prime}}
I_0\left [ 2\sqrt{y(z-z^\prime)} \right ]
J_1\left ( 2\sqrt{y z^\prime} \right )\\
F_{0,1}(y,z)&=&
e^{-2{\rm i}z}J_0\left ( 2\sqrt{y z} \right )\Theta( z)+
\Theta( z)\sqrt{y}\int_0^{ z} dz^\prime \frac{e^{-{\rm
i}2z^\prime}}{\sqrt{z-z^\prime}}
I_1\left [ 2\sqrt{y(z-z^\prime)} \right ]
J_0\left (2\sqrt{yz^\prime} \right )\\
F_{1,1}(y,z)&=&\Theta( z)\int_0^{ z} dz^\prime
e^{-{\rm i}2z^\prime}
I_0\left [ 2\sqrt{y(z-z^\prime)} \right ]
J_0\left ( 2\sqrt{y z^\prime} \right )\\
F_{2,0}(y,z)&=&\sqrt{\frac{z}{y}}I_1(2\sqrt{yz})\Theta(z)-
\Theta(z)\int_0^z{\rm d}z^\prime e^{-2{\rm i}z^\prime}\sqrt{\frac{z-z^\prime}{z^\prime}}
I_1[2\sqrt{y(z-z^\prime)}]J_1[2\sqrt{yz^\prime}]
\\
F_{0,2}(y,z)&=&e^{-2{\rm i}z}\sqrt{\frac{z}{y}}J_1(2\sqrt{yz})\Theta(z)+
\Theta(z)\int_0^{ z}{\rm d}z^\prime e^{-2{\rm i}z^\prime}\sqrt{\frac{z^\prime}{z-z^\prime}}
I_1[2\sqrt{y(z-z^\prime)}]J_1[2\sqrt{yz^\prime}]
\eea
\end{subequations}
with $J_i$ and $I_i$ the $i$th Bessel function of the first kind and the $i$th modified Bessel function respectively.

\section{Correlation Functions}
\bea \label{rho_jj}
G^{(1)}_{jj}(\xi_1,\xi_2;\tau)&=&\Gamma M(\xi_1,\xi_2)\int_{0}^{\Xi}d\xi'|\varphi(\xi^\prime)|^2 
F_{1,1}(\gamma_{\xi_2,\xi'},\tau)F_{1,1}^*(\gamma_{\xi_1,\xi'},\tau),\nonumber\\
&&+\delta_{j,+}M(\xi_1,\xi_2)
\int_{0}^{\tau}d\tau' 
F_{1,0}(\gamma_{\xi_1,0},\tau-\tau^\prime)F_{1,0}^*(\gamma_{\xi_2,0},\tau-\tau^\prime),\nonumber\\
\eea
where $M(\xi_1,\xi_2)=\Gamma \varphi(\xi_1)\varphi(\xi_2)$, $\delta_{j,j^\prime}$ is 
Kronecker's delta, and $\Xi={\min(\xi_1,\xi_2)}$.
\\
\begin{subequations}
\bea
G^{(2)}_{jj^\prime}(\xi_1,\xi_2;\tau)&=&\langle \hat{n}_{j}(\xi_1,\tau)\rangle 
\langle \hat{n}_{j^\prime}(\xi_2,\tau)\rangle+|\sigma_{jj^\prime}(\xi_1,\xi_2;\tau)|^2,\nonumber\\
\eea
where 
\be
\sigma_{jj}(\xi_1,\xi_2;\tau)=G^{(1)}_{jj}(\xi_1,\xi_2;\tau), 
\label{sigma_jj}
\ee
and 
\bea
\label{sigma_ij}
\sigma_{+-}(\xi_1,\xi_2;\tau)&\equiv& \langle\hat{\psi}_+(\xi_1,\tau)\hat{\psi}_-(\xi_2,\tau)\rangle\nonumber\\
&=&-M(\xi_1,\xi_2) F_{1,1}(\gamma_{\xi_2,\xi_1},\tau)\Theta(\xi_2-\xi_1)\\
&&-\Gamma M(\xi_1,\xi_2)\int_{0}^{\Xi}d\xi'|\varphi(\xi^\prime)|^2 
F_{1,1}(\gamma_{\xi_2,\xi'},\tau)F_{2,0}^*(\gamma_{\xi_1,\xi'},\tau),\nonumber
\eea
\end{subequations}

\end{appendix}


\begin{thebibliography}{99}
\bibitem{remark1} We neglect the external trapping potential 
and the atomic interactions, since they do not play a significant role on the time scales of 
the process under consideration.
\bibitem{remark3} The finite-width delta functions can be approximated by completely localized ones as they are integrated over slowly varying functions.
\end{thebibliography}
\end{document}